# Numerically Analytical Methods of Solution of the Diffraction Problem on the Nonlinear Dielectric Layer

Vasyl V. Yatsyk[*]

Usikov Institute for Radiophysics & Electronics
National Academy of Sciences of Ukraine

## Abstract

On an example of the open nonlinear electrodynamic system - transverse non-homogeneous, isotropic, nonlinear (a Kerr-like dielectric nonlinearity) dielectric layer, the algorithms of solution of the diffraction problem of a plane wave on the nonlinear object are shown. The first of them based on the iterative scheme is applicable to the solution of non-homogeneous system of the second kind nonlinear equations. The second algorithm bases on construction of system of the holomorphic nonlinear equations and application of the Newtonian method for the solution of the nonlinear equations system.

## 1 The diffraction problem on nonlinear dielectric layer

We consider a diffraction field (E-polarization case, $^{dif}_{(\alpha)}E_y = {}^{dif}_{(\alpha)}E_z = {}^{dif}_{(\alpha)}H_x = 0$) obtained by the incidence of a plane wave $^{inc}E_x(y,z) = {}^{inc}a\exp[i(\phi y - \Gamma \cdot (z - 2\pi\delta))]$, $z > 2\pi\delta$ on the transverse dielectric layer $\{(x,y,z): -\infty < x < \infty, -\infty < y < \infty, -2\pi\delta \le z \le 2\pi\delta\}$, which is non-homogeneous along the axis $0z$, homogeneous along the axis $0x$ and longitudinal direction $0y$, isotropic, nonlinear (a Kerr-like dielectric nonlinearity [1]), with the height $4\pi\delta$ and parameter of permittivity $_{(\alpha)}\varepsilon\left(z, \left|{}^{dif}_{(\alpha)}E_x\right|^2\right)$, see Fig. 1.

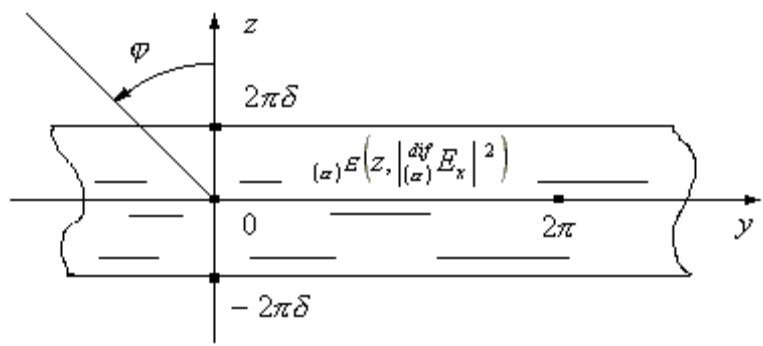

**Figure 1.** Nonlinear dielectric layer.

The complete diffraction field $^{dif}_{(\alpha)}E_x(y,z) = {}^{inc}E_x(y,z) + {}^{scat}_{(\alpha)}E_x(y,z)$ (here $^{scat}_{(\alpha)}E_x(y,z)$ is the scattering field) satisfies such conditions of the problem:

$$\left[\frac{\partial^2}{\partial y^2} + \frac{\partial^2}{\partial z^2} + \kappa^2 {}_{(\alpha)}\varepsilon\left(z, \left|{}^{dif}_{(\alpha)}E_x\right|^2\right)\right]{}^{dif}_{(\alpha)}E_x(y,z) = 0, \tag{1}$$

the generalized boundary conditions:

$$\overset{dif}{(\alpha)}E_{tg} \quad \text{and} \quad \overset{dif}{(\alpha)}H_{tg} \quad \text{are continuous at discontinuities} \quad {}_{(\alpha)}\varepsilon\left(z,\left|\overset{dif}{(\alpha)}E_x\right|^2\right);$$

$$\overset{dif}{(\alpha)}E_x(y,z)=\overset{dif}{(\alpha)}U(z)\cdot\exp(i\phi\,y), \text{ the condition of spatial quasihomogeneity on } y; \tag{2}$$

the condition of the radiation for scattered field:

$$\overset{scat}{(\alpha)}E_x(y,z)=\begin{Bmatrix}\overset{scat}{(\alpha)}a\\[2pt]\overset{scat}{(\alpha)}b\end{Bmatrix}\exp\left[i(\phi\,y\pm\Gamma(\kappa,\phi)(z\mp 2\pi\delta))\right], \quad z\overset{>}{<}\pm 2\pi\delta\,, \tag{3}$$

satisfying the requirement of absence of waves coming from infinity.

Here: $\overset{dif}{(\alpha)}E_x$, $\overset{dif}{(\alpha)}E_{tg}$, $\overset{dif}{(\alpha)}H_{tg}$ are the designated components of the field vectors $\vec{E}$ and $\vec{H}$; ${}_{(\alpha)}\varepsilon\left(z,\left|\overset{dif}{(\alpha)}E_x\right|^2\right)\equiv{}_{(\alpha)}\varepsilon\left(z,\left|\overset{dif}{(\alpha)}U(z)\right|^2\right)=\left\{\hat{\varepsilon}(z)+\alpha\left|\overset{dif}{(\alpha)}E_x\right|^2\equiv\hat{\varepsilon}(z)+\alpha\left|\overset{dif}{(\alpha)}U(z)\right|^2,\,|z|\leq 2\pi\delta,\,and\,1,\,|z|>2\pi\delta\right\}$ is the parameter of environment permittivity; $\alpha$ is the parameter of cubic electric susceptibility, $\operatorname{Re}\alpha\equiv const\geq 0$, $\operatorname{Im}\alpha\equiv 0$; $\hat{\varepsilon}(z)$ is the piecewise smooth function of variable $z$, $\operatorname{Re}\hat{\varepsilon}(z)>0$, $\operatorname{Im}\hat{\varepsilon}(z)\geq 0$; $\Gamma=(\kappa^2-\phi^2)^{1/2}$; $\exp(-i\omega t)$ is the time dependence; dimensionless parameters: $\omega=\kappa c$ is the circular frequency; $\kappa$ is the frequency $\kappa\in R$, describing the ratio of true height $h$ of a layer to the length $\lambda$ of the excitation wave, $h/\lambda=2\kappa\delta$; $c=(\varepsilon_0\,\mu_0)^{-1/2}$, $\operatorname{Im}c=0$, $\varepsilon_0$ and $\mu_0$ is the material parameters of environment; $\phi\equiv\kappa\cdot\sin(\varphi)\in R$, where $\varphi$ angle of fall of a plane wave ${}^{inc}E_x(y,z)$ which is counted in area $z>2\pi\delta$ from normal to a layer against a course of the hour arrow $|\varphi|<\pi/2$; amplitude of an incident diffraction field ${}^{inc}a$ are given.

The required solution of a problem (1)-(3) are of kind:

$$\overset{dif}{(\alpha)}E_x(y,z)=\overset{dif}{(\alpha)}U(z)\cdot e^{i\phi\,y}=\begin{cases}{}^{inc}a\cdot e^{i(\phi\,y-\Gamma\cdot(z-2\pi\delta))} \quad + \quad \overset{scat}{(\alpha)}a\cdot e^{i(\phi\,y+\Gamma\cdot(z-2\pi\delta))}, & z>2\pi\delta\,,\\[4pt]\overset{scat}{(\alpha)}U(z)\cdot e^{i\phi\,y}, & |z|\leq 2\pi\delta\,,\\[4pt]\overset{scat}{(\alpha)}b\cdot e^{i(\phi\,y-\Gamma\cdot(z+2\pi\delta))}, & z<-2\pi\delta\,.\end{cases} \tag{4}$$

Here $\overset{dif}{(\alpha)}U(-2\pi\delta)=\overset{scat}{(\alpha)}b$, $\overset{dif}{(\alpha)}U(2\pi\delta)={}^{inc}a+\overset{scat}{(\alpha)}a$.

The nonlinear diffraction problem (1)-(3) is reduced (following [2], [3]) to finding the solutions $\overset{dif}{(\alpha)}U(z)\in L_2([-2\pi\delta,2\pi\delta])$ of the non-homogeneous nonlinear integrated equation of the second kind:

$$\overset{dif}{(\alpha)}U(z)+\frac{i\kappa^2}{2\Gamma}\int\limits_{-2\pi\delta}^{2\pi\delta}\exp(i\,\Gamma\cdot|z-z_0|)\left[1-\left(\hat{\varepsilon}(z_0)+\alpha\left|\overset{dif}{(\alpha)}U(z_0)\right|^2\right)\right]\overset{dif}{(\alpha)}U(z_0)\,dz_0={}^{inc}U(z), \quad |z|\leq 2\pi\delta, \tag{5}$$

where ${}^{inc}U(z)={}^{inc}a\exp\left[-i\Gamma\cdot(z-2\pi\delta)\right]$.

## 2  Iteration scheme

The solution (5) with application of a quadrature method and use (4) is reduced to the non-homogeneous nonlinear equations system of the second kind:

$$(E - _{(\alpha)}B(\kappa, \phi, \left| _{(\alpha)}^{dif}U \right|^2)) \cdot _{(\alpha)}^{dif}U = {}^{inc}U \qquad (6)$$

Here $_{(\alpha)}^{dif}U = \{ _{(\alpha)}^{dif}U_n \}_{n=1}^N$ - vector-column of unknown $_{(\alpha)}^{dif}U_n = _{(\alpha)}^{dif}U(z_n, \kappa, \phi)$, given in units, $z_1 = -2\pi\delta < z_2 < ... < z_n < ... < z_N = 2\pi\delta$, $n = 1,2,...,N$; $N$ is the number of units, determining the order of system (6); $E = \{\delta_n^m\}_{n,m=1}^N$ (where $\delta_n^m$ is the Kronecker delta) and $_{(\alpha)}B\left(\kappa, \phi, \left| _{(\alpha)}^{dif}U \right|^2\right) = \left\{ A_n \cdot _{(\alpha)}K_{nm}\left(\kappa, \phi, \left| _{(\alpha)}^{dif}U \right|^2\right) \right\}_{n,m=1}^N$ is the matrix of dimension $N \times N$; $_{(\alpha)}K_{nm}\left(\kappa, \phi, \left| _{(\alpha)}^{dif}U \right|^2\right) = \frac{i \cdot \kappa^2}{2\Gamma(\kappa, \phi)} \exp\left(i \cdot \Gamma(\kappa, \phi) \cdot | z_n - z_m |\right)\left[1 - \left(\hat{\varepsilon}(z_m) + \alpha \left| _{(\alpha)}^{dif}U_m \right|^2\right)\right]$; $A_n$ are the numerical coefficients dictated by chosen quadrature form; the vector-column of the right hand part of (6) is the given current $^{inc}U = \left\{ ^{inc}U(z_n) \equiv {}^{inc}a \cdot \exp\left[-i \cdot \Gamma \cdot (z_n - 2\pi\delta)\right] \right\}_{n=1}^N$.

Solutions of non-homogeneous nonlinear system of the equations (6) are carried out by the method of iterations [4]. First step can be the finding of solution of the linear problem (case $\alpha = 0$). Each subsequent step of approximation represents the solution of linear diffraction problem received in result of linearization of the initial nonlinear system (6) for a nonlinear layer. This process can be presented as:

$$\left\{ \left( E - _{(\alpha)}B\left(\kappa, \phi, \left| _{(\alpha)}^{dif \ (s-1)}U \right|^2 \right) \right) \cdot _{(\alpha)}^{dif \ (s)}U = {}^{inc}U \right\}_{s=1}^{S \left\| _{(\alpha)}^{dif \ (s)}U - _{(\alpha)}^{dif \ (s-1)}U \right\| / \left\| _{(\alpha)}^{dif \ (s)}U \right\| < \xi_1} \qquad (7)$$

Here index $(s)$ is the designates the step of iteration, $S$ is the number of a step, $\xi_1$ is the given meaning of a relative error.

# 3 Newtonian algorithm

## 3. 1 Construction of system of the holomorphic nonlinear equations

Let us construct algorithm of the solution of the nonlinear integrated equation (5), which is based on Newtonian method of the solution of the nonlinear equation system. For this purpose we shall construct system of the nonlinear equations consisting of analytical functions.

Let's find the solution of the integrated equation (5) as:

$$_{(\alpha)}^{dif}U(z) = \sum_{n=0}^L {}_{(\alpha)}^{dif}c_n \cdot z^n = \sum_{n=0}^L \left( _{(\alpha)}^{dif}c_n' + i \cdot _{(\alpha)}^{dif}c_n'' \right) \cdot z^n; \quad _{(\alpha)}^{dif}c_n \in C, \ _{(\alpha)}^{dif}c_n', \ _{(\alpha)}^{dif}c_n'' \in R, \ |z| \le 2\pi\delta, (8)$$

where $_{(\alpha)}^{dif}c_n' = \text{Re}_{(\alpha)}^{dif}c_n$, $_{(\alpha)}^{dif}c_n'' = \text{Im}_{(\alpha)}^{dif}c_n$, parameter $L$ we choose from the assumption, that for the description $_{(\alpha)}^{dif}U(z)$ is enough to take $L + 1$ components in the expansion (8).

After substitution (8) in (5), calculations of value of the required function $_{(\alpha)}^{dif}U(z)$, and its derivatives $\left\{ \frac{\partial^l}{\partial z^l} \left( _{(\alpha)}^{dif}U(z) \right) \right\}_{l=1}^L$ in the point $z = 0$, we obtain the system of the nonlinear integrated equations concerning unknown $\left\{ _{(\alpha)}^{dif}c_n \right\}_{n=0}^L$:

$$\underset{(\alpha)}{\overset{dif}{}}c_l + \kappa^2 \frac{(-i\Gamma)^{l-1}}{\delta_l^0 + l} \int\limits_{-2\pi\delta}^{2\pi\delta} sign\left(z_0^{\ l}\right) e^{i\Gamma\cdot|z_0|} \left[1 - \left(\hat{\varepsilon}(z_0) + \alpha \left| \sum_{n=0}^{L} \underset{(\alpha)}{\overset{dif}{}}c_n z_0^{\ n} \right|^2 \right) \right] \left( \sum_{n=0}^{L} \underset{(\alpha)}{\overset{dif}{}}c_n z_0^{\ n} \right) dz_0 =$$

$$= \underset{}{\overset{inc}{}}a \cdot \frac{(-i\Gamma)^l}{\delta_l^0 + l} \cdot \exp(i\Gamma \cdot 2\pi\delta), \qquad\qquad l = 0, 1, 2, \dots, L \qquad (9)$$

Integrand functions in (9) are not holomorphic on $\left\{ \underset{(\alpha)}{\overset{dif}{}}c_n \right\}_{n=0}^{L}$. Let's write down (9) as system of differentiable functions in space of real values of arguments $\left\{ \underset{(\alpha)}{\overset{dif}{}}c_n', \ \underset{(\alpha)}{\overset{dif}{}}c_n'' \right\}_{n=0}^{L}$:

$$\begin{cases} \operatorname{Re} f_l\left( \underset{(\alpha)}{\overset{dif}{}}c_0', \ \underset{(\alpha)}{\overset{dif}{}}c_0'', \dots, \ \underset{(\alpha)}{\overset{dif}{}}c_L', \ \underset{(\alpha)}{\overset{dif}{}}c_L'' \right) = 0 \\ \operatorname{Im} f_l\left( \underset{(\alpha)}{\overset{dif}{}}c_0', \ \underset{(\alpha)}{\overset{dif}{}}c_0'', \dots, \ \underset{(\alpha)}{\overset{dif}{}}c_L', \ \underset{(\alpha)}{\overset{dif}{}}c_L'' \right) = 0 \end{cases}_{l=0}^{L}, \qquad \left\{ \underset{(\alpha)}{\overset{dif}{}}c_l', \ \underset{(\alpha)}{\overset{dif}{}}c_l'' \right\}_{l=0}^{L}, \left\{ \operatorname{Re} f_l, \operatorname{Im} f_l \right\}_{l=0}^{L} \in R, \quad (10)$$

where

$$f_l\left( \underset{(\alpha)}{\overset{dif}{}}c_0', \ \underset{(\alpha)}{\overset{dif}{}}c_0'', \dots, \ \underset{(\alpha)}{\overset{dif}{}}c_L', \ \underset{(\alpha)}{\overset{dif}{}}c_L'' \right) = \underset{(\alpha)}{\overset{dif}{}}c_l' + i \cdot \underset{(\alpha)}{\overset{dif}{}}c_l'' - \underset{}{\overset{inc}{}}a \cdot \frac{(-i \cdot \Gamma)^l}{\delta_l^0 + l} \cdot \exp(i\Gamma \cdot 2\pi\delta) + \kappa^2 \frac{(-i \cdot \Gamma)^{l-1}}{\delta_l^0 + l} \times$$

$$\times \int\limits_{-2\pi\delta}^{2\pi\delta} sign\left(z_0^{\ l}\right) \cdot e^{i\Gamma\cdot|z_0|} \left[ 1 - \hat{\varepsilon}(z_0) - \alpha \left( \left\{ \sum_{n=0}^{L} \underset{(\alpha)}{\overset{dif}{}}c_n' z_0^{\ n} \right\}^2 + \left\{ \sum_{n=0}^{L} \underset{(\alpha)}{\overset{dif}{}}c_n'' z_0^{\ n} \right\}^2 \right) \right] \left( \sum_{n=0}^{L} \left( \underset{(\alpha)}{\overset{dif}{}}c_n' + i \underset{(\alpha)}{\overset{dif}{}}c_n'' \right) z_0^{\ n} \right) dz_0 \qquad (11)$$

Let's write down system of $2L+2$ nonlinear equations (10) in a vector kind:

$$\vec{f}\left( \underset{(\alpha)}{\overset{dif}{}}\vec{c} \right) = \vec{0}, \qquad\qquad (12)$$

here: $\underset{(\alpha)}{\overset{dif}{}}\vec{c} \equiv \left( \underset{(\alpha)}{\overset{dif}{}}c_0', \ \underset{(\alpha)}{\overset{dif}{}}c_0'', \dots, \ \underset{(\alpha)}{\overset{dif}{}}c_L', \ \underset{(\alpha)}{\overset{dif}{}}c_L'' \right)^T$ - column vector of required solution, $\vec{f}\left( \underset{(\alpha)}{\overset{dif}{}}\vec{c} \right) \equiv \left( \operatorname{Re} f_0\left( \underset{(\alpha)}{\overset{dif}{}}\vec{c} \right), \operatorname{Im} f_0\left( \underset{(\alpha)}{\overset{dif}{}}\vec{c} \right), \dots, \operatorname{Re} f_L\left( \underset{(\alpha)}{\overset{dif}{}}\vec{c} \right), \operatorname{Im} f_L\left( \underset{(\alpha)}{\overset{dif}{}}\vec{c} \right) \right)^T$ - column vector of functions, $T$ - transposition.

## 3. 2 Explicit form of the nonlinear equations system and the Jacobi matrixes. Newtonian method

The solution (12) could be found by the method of Newton for system of the nonlinear equations [5]:

$$J^k \cdot \left( \underset{(\alpha)}{\overset{dif}{}}\vec{c}^{\ k+1} - \underset{(\alpha)}{\overset{dif}{}}\vec{c}^{\ k} \right) = -\vec{f}\left( \underset{(\alpha)}{\overset{dif}{}}\vec{c}^{\ k} \right), \qquad \text{when } k \to \infty, \ \underset{(\alpha)}{\overset{dif}{}}\vec{c}^{\ k} \to \underset{(\alpha)}{\overset{dif}{}}\vec{c} \qquad (13)$$

where $J^k \equiv J^k\left( \underset{(\alpha)}{\overset{dif}{}}\vec{c}^{\ k} \right)$ is the $k$-th iteration of the Jacobi matrixes $J\left( \underset{(\alpha)}{\overset{dif}{}}\vec{c} \right)$ for $\vec{f}\left( \underset{(\alpha)}{\overset{dif}{}}\vec{c} \right)$. Initial approach for the process (13) may be taken from the solution of the appropriate linear problem (see (12) for $\alpha = 0$).

Let the linear part of dielectric permeability is given as:

$$\hat{\varepsilon}(z) = \sum_{g=0}^{G} \varepsilon_g z^g, \qquad\qquad \text{here } \left\{ \varepsilon_g \right\}_{g=0}^{G} \in C, \ |z| \le 2\pi\delta, \qquad (14)$$

then for (13) the vector function and Jacobi matrix have the form:

$$
\vec{f}\begin{pmatrix}dif\\(\alpha)\vec{c}\end{pmatrix} \equiv
\begin{pmatrix}
\operatorname{Re} f_0\begin{pmatrix}dif\\(\alpha)\vec{c}\end{pmatrix}\\
\operatorname{Im} f_0\begin{pmatrix}dif\\(\alpha)\vec{c}\end{pmatrix}\\
\dots\dots\dots\\
\dots\dots\dots\\
\dots\dots\dots\\
\dots\dots\dots\\
\operatorname{Re} f_L\begin{pmatrix}dif\\(\alpha)\vec{c}\end{pmatrix}\\
\operatorname{Im} f_L\begin{pmatrix}dif\\(\alpha)\vec{c}\end{pmatrix}
\end{pmatrix},\;
J\begin{pmatrix}dif\\(\alpha)\vec{c}\end{pmatrix} \equiv
\begin{pmatrix}
\operatorname{Re}\dfrac{\partial f_0\begin{pmatrix}dif\\(\alpha)\vec{c}\end{pmatrix}}{\partial_{(\alpha)}^{dif}c_0'}, \operatorname{Re}\dfrac{\partial f_0\begin{pmatrix}dif\\(\alpha)\vec{c}\end{pmatrix}}{\partial_{(\alpha)}^{dif}c_0''},\dots,\operatorname{Re}\dfrac{\partial f_0\begin{pmatrix}dif\\(\alpha)\vec{c}\end{pmatrix}}{\partial_{(\alpha)}^{dif}c_L'}, \operatorname{Re}\dfrac{\partial f_0\begin{pmatrix}dif\\(\alpha)\vec{c}\end{pmatrix}}{\partial_{(\alpha)}^{dif}c_L''}\\
\operatorname{Im}\dfrac{\partial f_0\begin{pmatrix}dif\\(\alpha)\vec{c}\end{pmatrix}}{\partial_{(\alpha)}^{dif}c_0'}, \operatorname{Im}\dfrac{\partial f_0\begin{pmatrix}dif\\(\alpha)\vec{c}\end{pmatrix}}{\partial_{(\alpha)}^{dif}c_0''},\dots,\operatorname{Im}\dfrac{\partial f_0\begin{pmatrix}dif\\(\alpha)\vec{c}\end{pmatrix}}{\partial_{(\alpha)}^{dif}c_L'}, \operatorname{Im}\dfrac{\partial f_0\begin{pmatrix}dif\\(\alpha)\vec{c}\end{pmatrix}}{\partial_{(\alpha)}^{dif}c_L''}\\
\dots\dots\dots\dots\dots\dots\dots\dots\dots\dots\dots\dots\dots\dots\dots\dots\\
\operatorname{Re}\dfrac{\partial f_L\begin{pmatrix}dif\\(\alpha)\vec{c}\end{pmatrix}}{\partial_{(\alpha)}^{dif}c_0'}, \operatorname{Re}\dfrac{\partial f_L\begin{pmatrix}dif\\(\alpha)\vec{c}\end{pmatrix}}{\partial_{(\alpha)}^{dif}c_0''},\dots,\operatorname{Re}\dfrac{\partial f_L\begin{pmatrix}dif\\(\alpha)\vec{c}\end{pmatrix}}{\partial_{(\alpha)}^{dif}c_L'}, \operatorname{Re}\dfrac{\partial f_L\begin{pmatrix}dif\\(\alpha)\vec{c}\end{pmatrix}}{\partial_{(\alpha)}^{dif}c_L''}\\
\operatorname{Im}\dfrac{\partial f_L\begin{pmatrix}dif\\(\alpha)\vec{c}\end{pmatrix}}{\partial_{(\alpha)}^{dif}c_0'}, \operatorname{Im}\dfrac{\partial f_L\begin{pmatrix}dif\\(\alpha)\vec{c}\end{pmatrix}}{\partial_{(\alpha)}^{dif}c_0''},\dots,\operatorname{Im}\dfrac{\partial f_L\begin{pmatrix}dif\\(\alpha)\vec{c}\end{pmatrix}}{\partial_{(\alpha)}^{dif}c_L'}, \operatorname{Im}\dfrac{\partial f_L\begin{pmatrix}dif\\(\alpha)\vec{c}\end{pmatrix}}{\partial_{(\alpha)}^{dif}c_L''}
\end{pmatrix}
\tag{15}
$$

at the following functions (for $l, p = 0, 1, 2, \dots, L$), see (11):

$$
f_l\begin{pmatrix}dif\\(\alpha)\vec{c}\end{pmatrix} = {}_{(\alpha)}^{dif}c_l' + i\,{}_{(\alpha)}^{dif}c_l'' - {}^{inc}a\,\frac{(-i\Gamma)^l}{\delta_l^0 + l}\cdot \exp\left(i\Gamma\,2\pi\delta\right) + \kappa^2\,\frac{(-i\Gamma)^{l-1}}{\delta_l^0 + l}\sum_{s=0}^{L}\left({}_{(\alpha)}^{dif}c_s' + i\cdot{}_{(\alpha)}^{dif}c_s''\right)\times
$$
$$
\times\left\{V(l,s) - \sum_{g=0}^{G}\varepsilon_g V(l,s+g) - \alpha\sum_{n=0}^{L}\sum_{m=0}^{L}\left({}_{(\alpha)}^{dif}c_n'\cdot{}_{(\alpha)}^{dif}c_m' + {}_{(\alpha)}^{dif}c_n''\cdot{}_{(\alpha)}^{dif}c_m''\right)\cdot V(l,n+m+s)\right\};
$$

$$
\frac{\partial f_l\begin{pmatrix}dif\\(\alpha)\vec{c}\end{pmatrix}}{\partial_{(\alpha)}^{dif}c_p'} = \delta_p^l + \kappa^2\,\frac{(-i\cdot\Gamma)^{l-1}}{\delta_l^0 + l}\cdot\Big[V(l,p) - \sum_{g=0}^{G}\varepsilon_g\cdot V(l,p+g) -
$$
$$
- \alpha\sum_{n=0}^{L}\sum_{m=0}^{L}\left(3\,{}_{(\alpha)}^{dif}c_n'\cdot{}_{(\alpha)}^{dif}c_m' + {}_{(\alpha)}^{dif}c_n''\cdot{}_{(\alpha)}^{dif}c_m'' + 2i\,{}_{(\alpha)}^{dif}c_n'\cdot{}_{(\alpha)}^{dif}c_m''\right)\cdot V(l,n+m+p)\Big];
$$

$$
\frac{\partial f_l\begin{pmatrix}dif\\(\alpha)\vec{c}\end{pmatrix}}{\partial_{(\alpha)}^{dif}c_p''} = i\cdot\delta_p^l + i\cdot\kappa^2\,\frac{(-i\cdot\Gamma)^{l-1}}{\delta_l^0 + l}\cdot\Big[V(l,p) - \sum_{g=0}^{G}\varepsilon_g\cdot V(l,p+g) -
$$
$$
- \alpha\sum_{n=0}^{L}\sum_{m=0}^{L}\left({}_{(\alpha)}^{dif}c_n'\cdot{}_{(\alpha)}^{dif}c_m' + 3\,{}_{(\alpha)}^{dif}c_n''\cdot{}_{(\alpha)}^{dif}c_m'' - 2i\,{}_{(\alpha)}^{dif}c_n'\cdot{}_{(\alpha)}^{dif}c_m''\right)\cdot V(l,n+m+p)\Big].
$$

(16)

Here $V(l,p) \equiv \int\limits_{-2\pi\delta}^{2\pi\delta} sign(z_0^l)\cdot\exp\left(i\cdot\Gamma\cdot|z_0|\right)\cdot z_0^p\,dz_0 = \left(1 + (-1)^{l+p}\right)\cdot W(p)$ and $W(p) = \exp\left(i\cdot\Gamma\cdot2\pi\delta\right)\times$

$$
\times\left[\frac{(2\pi\delta)^p}{i\cdot\Gamma} + \left(1 - \delta_p^0\right)\sum_{k=1}^{p}(-1)^k\,\frac{p(p-1)\cdot\dots\cdot(p-k+1)}{(i\cdot\Gamma)^{k+1}}(2\pi\delta)^{p-k}\right] - \frac{\delta_p^0}{i\cdot\Gamma} - \left(1 - \delta_p^0\right)(-1)^p\,\frac{p(p-1)\cdot\dots\cdot1}{(i\cdot\Gamma)^{p+1}}.
$$

## 4    Conclusion

The submitted algorithms have the following features. The first, iterative algorithm (7) may be referred to direct numerical methods. It is simple in realization, but has lacks inherent in the iterative circuits based on quadrature forms. Second algorithm (see (8), (12)-(16)) on the basis of the Newtonian method (13). It is numerically analytical with has good (quadric) convergence. It uses advantages of analytical approaches. This approach consists from: construction of system of the holomorphic nonlinear equations; obtain of explicit form of the nonlinear equations system and the Jacobi matrix; numerical realization of the method of Newton for nonlinear systems of the equations and to receive analytical representation of the solution of a nonlinear problem of diffraction. The proposed algorithms (7) and (13) for the solution of nonlinear diffraction problem are applied: at investigation of processes of wave self-influence [1]; at the analysis of amplitude-phase dispersion of eigen oscillation-wave fields in the nonlinear objects [4]; at development of approaches of numerically analytical description of the equations of dispersion and space-time evolutions of field of nonlinear electrodynamic structure near Morse critical point [6].